# The Curious Case of Max Planck's "retracted" papers.
## When past scientific practices meet contemporary publishing norms

Yves Gingras (gingras.yves@uqam.ca), Université du Québec à Montréal
Mahdi Khelfaoui (mahdi.khelfaoui@uqtr.ca), Université du Québec à Trois-Rivières

**Abstract**

This article examines the case of two papers published in Naturwissenschaften by the physicist Max Planck that were retrospectively marked as "retracted" on Springer's digital platform. Rather than originating in scientific fraud, these withdrawals appear to result from contemporary digitization and copyright-management procedures applied anachronistically to historical publications. Through an investigation of the circulation history of Planck's 1940 and 1942 philosophical essays, the article shows that republication across multiple formats was a common and legitimate practice within the scientific publishing culture of the early 20th century. Such practices only became problematic with the later transformation of the scientific article into a countable and proprietary unit within systems of bibliometric evaluation and commercial academic publishing. This article argues that contemporary notions such as "duplicate publication" and "self-plagiarism" are historically situated categories that cannot be applied retrospectively without distorting the historical record. More broadly, the Planck case reveals how digital scholarly infrastructures controlled by large commercial publishers can limit the accessibility of the scientific past.

While browsing the Retraction Watch website, we came across a piece titled "Retractions by Nobel Prize Winners" (Retraction Watch, n.d.) and, as historians of physics, we were very surprised to find Max Planck's name in that list. Two of his papers, published in the leading German scientific journal *Die Naturwissenschaften* in 1940 and 1942 (Planck 1940, 1942a), were marked as "retracted" on the journal's online platform, owned by Springer since its creation in 1913. The fact that no clear explanation was provided and that the corresponding pages had been left blank, was intriguing. As the historian John Heilbron wrote in his major biography of the German physicist, Planck was an "upright man", founder of quantum theory in 1900 and recipient of the 1918 Nobel Prize in Physics (Heilbron 1986). At the time these two now "retracted" papers were published, he was one of the most renowned living physicists and, as the elder "spokesman for German Science", to use Heilbron's formula, was often invited to give talks reflecting on the nature of the discipline.

We therefore found it difficult to believe that these papers could really have been retracted during his lifetime (Planck died in 1947) or even that there were good reasons to retract them later. Rather, we suspected that we were dealing with a recent and anachronistic decision of the journal's publisher, based on a misunderstanding, or ignorance, of past publication practices. This article thus investigates why these two papers came to be retrospectively classified as "retracted" within Springer's digital publishing system. More generally, this case offers an opportunity to reflect on the historical transformation of scientific publication practices and on the problems that arise when contemporary categories such as "duplicate publication", "copyright violation", or "self-plagiarism" are retrospectively applied to texts produced within very different epistemic and editorial contexts. While these two papers could be considered as anomalous events in the history of scholarly publishing, we rather argue that they are inadvertent consequences of the broader historical transformation of the scientific article into an accounting "unit" of knowledge within contemporary systems of academic evaluation, as well as its transformation into a proprietary and profitable product in the hands of major private publishers.

**The rise of retraction tracking and hunting**

The growing interest in questions of scientific integrity since the 1980s first led to the creation of federal institutions dedicated to research misconduct oversight in the United States, eventually leading to the establishment of the Office of Research Integrity in 1992 (Price 2013). The Internet revolution that followed, however, made possible a much larger participation of scientists in the identification of problematic papers. This trend contributed to the creation in 2010 of the platform *Retraction Watch*, now a widely known and very useful website dedicated to tracking and discussing scientific retractions and, more generally, issues related to research integrity. Emerging at a time when both the scientific community and the broader public were increasingly concerned about a perceived rise in research misconduct, the platform now documents several thousand retracted papers and problematic cases involving plagiarism, duplicate publication, data fabrication, and image manipulation. Over time, it has become the principal database through which retracted scientific articles are registered (Oransky 2020). Over the past two decades, platforms such as *Retraction Watch* have contributed to the institutionalization of retraction tracking. Retractions, whether initiated by authors or by journal editors, are now continuously indexed, monitored, and debated on various online platforms devoted to scientific integrity.

Another popular website where these discussions take place is PubPeer, created in 2012 to facilitate post-publication commentary on scientific papers (Barbour and Stell 2020). Not surprisingly, some participants on the platform flagged Planck's two papers in 2023 after noticing that they had been marked as "retracted" on the journal's electronic platform (PubPeer 2023). Since no date accompanies the withdrawal notice, determining precisely when these "retractions" where decided is difficult. Nevertheless, metadata retrieved through the Crossref API (https://api.crossref.org/works?filter=updates:10.1007/bf01488952 and

https://api.crossref.org/works?filter=updates:10.1007/bf01475382) indicate that the DOI records associated with the two papers were created in April 2005.[1]

Springer's own presentation (https://preview.springer.com/it/about-springer/history) of its online journal archives similarly indicates that the large-scale digitization of older scientific periodicals largely took place around the turn of the 21st century, when publishers progressively integrated historical collections into searchable digital platforms, thus adding more value to their product. It therefore seems plausible that the decision to "retract" Planck's papers was made at that time and applied rather mechanically by Springer according to some internal cataloguing criteria, with little attention paid to the historical context of the publications themselves. It is somewhat ironic that the few comments posted on PubPeer treated the two papers as ordinary cases of "retraction", one commentator even asking whether Planck had agreed to retract them, as if publication norms and practices had remained unchanged since the beginning of modern science. As we will see later, this is far from the case. Before examining publication practices at the time of Planck, let us first try to identify the actual reasons behind these curious cases of retraction.

**Investigating the withdrawal notices**

As is well known, most scientific journals abandoned print publication and moved to electronic platforms at the turn of the 21st century. Aware of the continuing value of older printed editions, publishers have, over time, integrated digitized PDF versions of earlier volumes into their online archives. The PDF versions of Planck's two retracted papers carry the statement "This article has been withdrawn due to article violation", a curious wording that does not really make sense, while Springer's website displays the more meaningful notice: "This article has been withdrawn due to copyright violation". This makes clear that the issue had less to do with the scientific validity of the two papers than with questions related to copyright ownership at the time the journal was digitized and made available on Springer's platform.

The text of a retracted scientific article usually remains accessible in its original form alongside a withdrawal notice, but remarkably, the two PDF files available on *Naturwissenschaften*'s digital platform contain only blank pages where the articles originally appeared: two (778-779) for the 1940 paper and nine pages (125-133) for the 1942 paper (Figure 1). All this clearly suggests that some lawyer at Springer was overshadowing the process and considered these papers as problematic forms of "duplicate publications". Thankfully, both Planck Papers can still be read through the nonprofit digital library Internet Archive, which hosts scanned versions of *Naturwissenschaften* volumes from 1913 to 2001

[1] Thanks to our colleague Guillaume Cabanac for bringing that reference to Crossref to our attention.

(Figure 2). It is important to note that these two papers by Planck are not standard research articles reporting new experimental or theoretical results, but philosophical reflections on the nature of scientific knowledge. To understand why texts of this kind were published in *Naturwissenschaften*, let us briefly examine the editorial culture of the journal at that time.

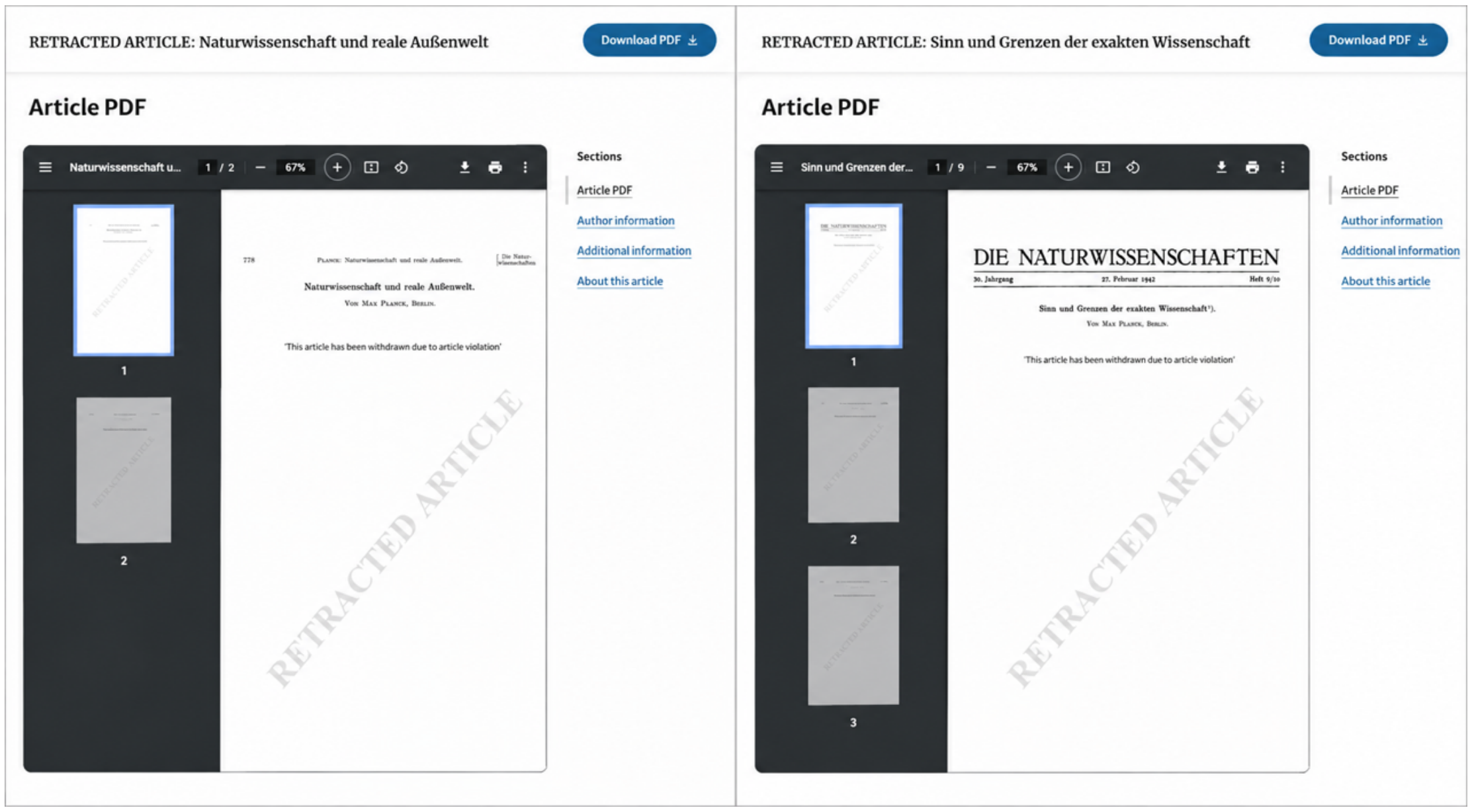

**Figure 1 - Screenshots of the Springer platform displaying Max Planck's 1940 and 1942 papers as "retracted"**

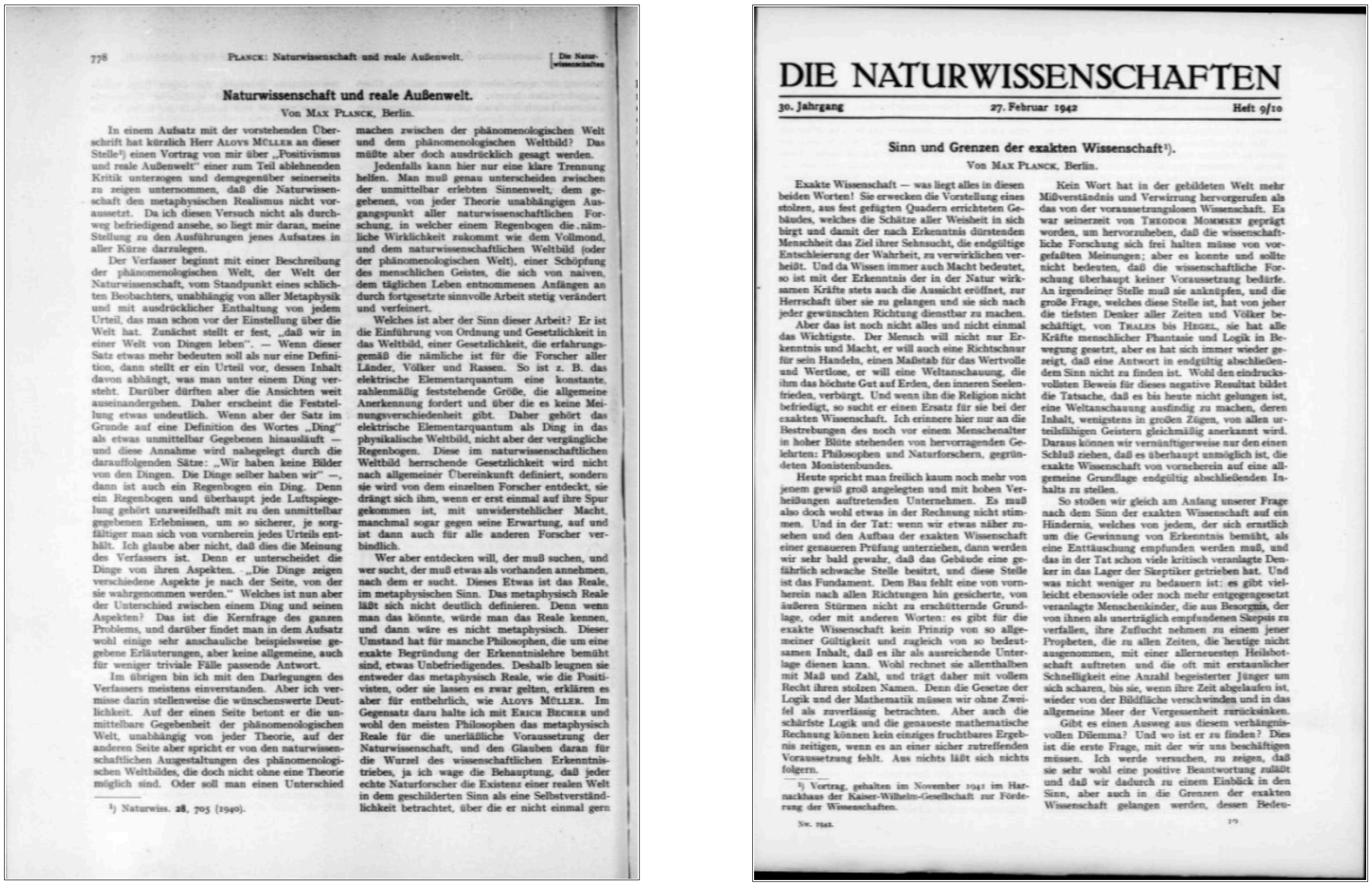
778 Planck: Naturwissenschaft und reale Außenwelt. Die Naturwissenschaften

Naturwissenschaft und reale Außenwelt.

Von Max Planck, Berlin.

[1] Naturwiss. 28, 705 (1940).

DIE NATURWISSENSCHAFTEN

30. Jahrgang 27. Februar 1942 Heft 9/10

Sinn und Grenzen der exakten Wissenschaft[1].

Von Max Planck, Berlin.

[1] Vortrag, gehalten im November 1941 im Harnackhaus der Kaiser-Wilhelm-Gesellschaft zur Förderung der Wissenschaften.

**Figure 2 - Scanned versions of Max Planck's 1940 and 1942 papers available through Internet Archive**

Founded in 1913 by the physicist Arnold Berliner and published by Julius Springer Verlag, *Naturwissenschaften* was conceived as a German counterpart to the British scientific magazine *Nature*. As Berliner explained in his correspondence with the publisher before the launch of the journal, the objective was not simply to create another specialized scientific periodical, but a broad scientific weekly capable of informing researchers about developments occurring among the natural and medical sciences (Thatje 2013). The original subtitle of the journal, *Wochenschrift für die Fortschritte der Naturwissenschaften, der Medizin und der Technik* ("Weekly journal for the advances of the natural sciences, medicine, and technology"), reflected this ambition. Under Berliner's editorship, *Naturwissenschaften* rapidly became one of the central German-speaking scientific magazine and published contributions by many of the leading physicists of the early twentieth century, including Albert Einstein, Max Planck, Niels Bohr, Werner Heisenberg, and Erwin Schrödinger (Autrum 1988). Unlike the specialized disciplinary journals that came to dominate scientific publishing after the Second World War (Khelfaoui and Gingras 2019), *Naturwissenschaften*, like *Nature*, occupied an intermediate position between a research journal, an intellectual forum, and vehicle for scientific communication across disciplines (Lamla 1963). Alongside technical papers, the journal regularly published lectures, conference reports, and discussions concerning the philosophical and cultural implications of scientific developments. This editorial culture helps explain why contributions of philosophical nature such as Planck's 1940 and 1942 essays were welcomed.

Let us first examine the article "Sinn und Grenzen der exakten Wissenschaft" ("Meaning and Limits of Exact Science") which appeared in *Naturwissenschaften* in 1942. The text circulated in several forms between 1941 and 1943. Originally delivered as a lecture at the Kaiser-Wilhelm-Gesellschaft in Berlin in November 1941, it was published the following year as a booklet in Leipzig by the editor Johann Ambrosius Barth (Planck 1942c), but also as an article in the intellectual journal *Europäische Revue* (Planck 1942b), as well as in *Naturwissenschaften*. In 1943, the text was also incorporated into an expanded two-volume edition of *Wege zur physikalischen Erkenntnis. Reden und Vorträge von Max Planck,* a collection of essays and lectures originally published in Leipzig in 1933 (Planck 1943).

Unlike the 1942 paper, which has been printed in many venues that may explain its recent "retraction", the case for having also retracted the 1940 paper "Naturwissenschaft und reale Außenwelt" ("Natural Science and the Real External World") is more mysterious. We have found no evidence that this short two-page paper circulated in printed form either before or after its publication in *Naturwissenschaften*. Nor does it appear to have circulated through translations or parallel publications. The paper in fact belongs to Planck's long-standing critique of positivism. Earlier that same year, Aloys Müller had published in *Naturwissenschaften* a paper bearing the exact same title (Müller 1940), in which he critically discussed Planck's philosophical position while referring to his 1931 essay *Positivismus und reale Außenwelt*, itself derived from a conference he had given in November 1930 at the Kaiser Wilhelm-Gesellschaft (Planck 1931). Planck then responded to Müller in the same journal a few months later using the identical title "Naturwissenschaft und reale Außenwelt". This text is thus a short philosophical response within an ongoing intellectual debate. While it briefly revisits themes Planck had developed in the 1931 booklet of thirty-five pages, it is of course far from being a reproduction of that earlier essay.

So, why are Planck's two papers now invisible on the contemporary digital version of *Naturwissenschaften*? The most plausible explanation is that the decision to "retract" them, and to even erase their texts from the journal's electronic archive, resulted from Springer's modern digitization and rights-management procedures. In the case of the 1942 paper, the withdrawal could be explained by the multiple republications of that lecture before and after its appearance in *Naturwissenschaften,* a situation that would have been interpreted retrospectively as violating modern copyright and reproduction norms during the digitization process. Particularly significant is the fact that Springer itself republished the 1941 lecture in 2001 in the collection *Vorträge Reden Erinnerungen,* a volume gathering Planck's philosophical essays and lectures on science (Planck 2001). During the digitization of *Naturwissenschaften*, possibly shortly after this republication, the coexistence of the 1942 journal article and the later book chapter carrying the identical title, "Sinn und Grenzen der exakten Wissenschaft", may have generated confusion within Springer's rights-management and cataloguing systems.

The case of the 1940 paper is more puzzling since, as we have seen, no other printed version appears to exist. One plausible explanation for the decision to withdraw that paper is that it also resulted from a cataloguing ambiguity created by the existence, in the same journal and only a few months apart, of two different papers by different authors (Aloys Müller and Max Planck) bearing the exact same title, "Naturwissenschaft und reale Außenwelt". The reuse of Müller's title by Planck was of course likely intended to signal direct participation in an ongoing philosophical debate. While such a situation was not problematic within the publishing culture of the time, it may have generated confusion within contemporary digital indexing and rights-management systems.

**Putting "duplicate publications" in their historical context**

Viewed from today's publishing norms, the multiple circulation of Planck's 1942 article can easily be interpreted as a case of "duplicate publication" and condemned as what is now commonly known as "self-plagiarism". Yet, applying such a category retrospectively raises historical problems, since "self-plagiarism" is a relatively recent notion that emerged alongside the rise of bibliometric evaluation of researchers from the 1990s onward (Cronin 2013; Horbach and Halffman 2019; Gingras 2023). In such evaluation systems, scientific papers came to function as "accounting" units of academic "productivity", making the republication of substantially similar texts appear problematic and unethical (Andreescu 2013; Gingras 2020). The emergence of electronic text-detection tools further reinforced these concerns by making textual overlap more easily detectable and by facilitating the automated identification of similar passages across scientific publications (Andreescu 2013). Thus "self-plagiarism" should not only be understood as an ethical category, but also as a bureaucratic and bibliometric one, closely tied to systems of academic evaluation that are based on publication counts and quantitative performance metrics not always congruent with the maximization of knowledge circulation.

Such an obsession with publication productivity, however, did not exist in Planck's time. Scholarly communication operated according to rather different assumptions, among them the importance of furthering the largest possible dissemination of knowledge across fragmented linguistic and national scientific communities at a time when rapid access to foreign publications remained limited. In the first half of the twentieth century, republication across multiple journals, languages, or audiences was therefore not considered a breach of originality or copyright but was instead understood as a legitimate means of extending the circulation of scientific ideas. It was then not uncommon to see a paper published in English in *Philosophical Magazine* also appear in German in *Zeitschrift für physikalische Chemie* or in *Physikalische Zeitschrift*. Prestigious lectures, such as those of Planck, frequently circulated simultaneously as conference proceedings, journal or magazine articles, booklets, or collected essays. The boundaries separating these formats were far more porous than they are today.

Niels Bohr's lecture "The Quantum Postulate and the Recent Development of Atomic Theory", which introduced the principle of complementarity and became one of the foundational texts of the Copenhagen interpretation of quantum mechanics, offers a good example. First delivered at the Como Conference in September 1927, the lecture subsequently appeared in the conference proceedings, but was also published in English in *Nature* in 1928 and translated into German that same year for publication in *Naturwissenschaften*. Similar patterns can be observed in the case of Henri Poincaré's 1904 lecture on the principle of relativity. First delivered at the International Congress of Arts and Science in Saint Louis, the lecture circulated through several publication channels. It appeared in *La Revue des idées* in November 1904, was republished shortly afterward in the specialized *Bulletin des sciences mathématiques* and translated into English in *The Monist* in January 1905 (Giacomini 2026).

These examples, and there are many, illustrate a publishing culture in which scientific articles were not treated as quantifiable units within systems of academic evaluation, nor constrained by strong proprietary conceptions of “copyright” (Patterson 1968; Fyfe et al. 2022). As Alex Csiszar has shown, the scientific article progressively emerged during the $19^{th}$ century as a stable textual unit associated with scientific authority and priority claims (Csiszar 2018). Yet the academic publishing culture of the early $20^{th}$ century tolerated considerable fluidity between lectures, proceedings, journal articles, translations, and collected essays, especially for reflective or synthetic texts such as those published by Planck, Bohr, or Einstein. Their republication and multilingual circulation were generally considered as normal mechanisms for disseminating scientific ideas across different intellectual and national audiences.

However, this emphasis on circulation should not be taken to mean that early twentieth-century scientific publishing operated outside commercial logics. As already mentioned, *Naturwissenschaften* was itself a commercial enterprise published by Julius Springer Verlag, and commercial publishers had long played an important role in the organization and circulation of scientific knowledge. Scholarly and commercial logics therefore coexisted throughout the history of scientific publishing. Yet, although no journal aimed to lose money, they clearly understood that diffusion was more important than a strict application of intellectual property rights, which in fact were rarely explicitly signed off by authors. When republication happened, journals would at most ask that the original place of publication be mentioned. What distinguished much of the first half of the twentieth century (not to mention the earlier period) was the fact that commercial interests remained largely subordinated to, or constrained by, the norms of scholarly communication, which aimed at maximising diffusion. So even where copyright existed, it was rarely aggressively enforced.

This situation changed progressively after the second half of the twentieth century, as scientific journals increasingly came under the control of a small number of large commercial publishing firms operating according to market logics very different from those that had historically characterized learned societies and university presses. The postwar expansion of universities and international research communities transformed academic publishing into a highly profitable sector structured around subscription markets and the management of publishing rights (Fyfe et al. 2022), and toward the end of the century, the balance between scholarly and commercial interests had considerably shifted. Moreover, scientific journals acquired a new institutional role within academic life, since scientific articles and journals gradually became instruments through which universities evaluated researchers. Publication records acquired a bureaucratic importance and became closely tied to hiring, promotion, and funding decisions, while publishers progressively reinforced their legal control over scientific texts through formal copyright transfer agreements and, later, digital rights management.

## When anachronistic norms erase historical knowledge

Viewed from a historical perspective, it is obvious that Planck's publication practices were not considered problematic and in need of any kind of "retraction". Applying modern definitions of "copyright" and "duplicate publication" to historical sources is highly problematic, since it effectively rewrites the historical record by making important sources invisible on the digital platform of the journal in which they were first printed. The privatization and commercialization of journals (Larivière, Haustein and Mongeon 2015; Khelfaoui and Gingras 2020) whose original mission was based on a communal understanding of scientific knowledge and its circulation has thus led to a reinterpretation of the status of older publications through the lens of the capitalist notion of private property under "copyright". Not only were Planck's papers never retracted before the digitalization of *Naturwissenschaften*, but they also continued to be cited normally in the scholarly literature well into the twenty-first century. Heilbron's biography refers to the 1940 paper, while the 1942 paper was still being cited as late in 2013 in the *Journal of the History of Biology* (Rieppel, Williams and Ebach 2013). In none of these cases were the papers described as "retracted", especially since they were essentially philosophical lectures and reflections, and did not claim to offer new empirical facts or theories.

Our investigation led us to conclude that the "retraction" of Max Planck's two articles did not originate in the epistemic practices of the scientific community of his time but was a byproduct of the contemporary norms of digital scholarly platforms, which are dominated by large publishing groups increasingly sensitive to copyright questions tied to commercialization and profit. The growing control of scientific publishing by these groups has also contributed to make the scientific article a distinct and proprietary textual object. Earlier forms of scientific communication tolerated a much greater fluid circulation between printed media. Lectures, booklets, conference proceedings, journal articles and their translations frequently represented complementary forms of circulation of the same intellectual content aimed at different audiences. By contrast, contemporary digital publishing systems rely on the opposite assumption. They require clearly individualized and legally identifiable publication units associated with precise ownership, metadata and citation practices.

In that sense, retracting Max Planck's papers based on contemporary notions of copyright is not only anachronistic but, more importantly, obscures important historical sources from view. The legal status of many of older scientific publications also remains historically ambiguous. Max Planck died in 1947, meaning that, in several jurisdictions, his writings have now entered the public domain. Moreover, publishing agreements signed in the 1930s and 1940s obviously did not anticipate contemporary forms of digital dissemination through online platforms and globally searchable PDF archives. As a result, access to older scientific literature now depends on large commercial digital infrastructures such as SpringerLink, Wiley Online Library, or Elsevier's databases, which now mediate the visibility and accessibility of a substantial part of the scientific past. In that context, the strong control now exercised by commercial conglomerates over the digital circulation of such historical texts is much more problematic than current publishing practices suggest.

What first seemed to be a curious case of retraction involving a famous Nobel prize of physics finally appears as an instance of arbitrary and anachronistic decisions made by the contemporary owners of a historically important scientific journal, *Naturwissenschaften* rebranded as *The Science of Nature* since 2013 (Khelfaoui and Gingras 2025). The case further illustrates the considerable power now exercised by digital publishing infrastructures over the accessibility and classification of the scientific past. Historians of science should not accept that historical sources be erased in the name of problematic contemporary notions of legitimacy and should from now also pay more attention to the contemporary editorial and technical interventions through which historical scientific publications are made digitally accessible and classified.

**Epilogue**

Shortly after the first version of this article was posted on arXiv in May 2026, the case attracted the attention of *Science* (Kean 2026) and was subsequently covered extensively by online media and newspapers in several languages and countries, including Germany, Sweden, the United States, and Brazil (Schnabel 2026; Glad 2026; Ouellette 2026; Leite 2026). Curiously, *Nature* has remained silent on the case, despite its News section being supposedly "editorially independent" of its publisher, Springer Nature. Confronted with such media frenzy and repeated requests to explain these bizarre retractions, Springer Nature launched an internal investigation and revealed that the withdrawals had occurred on 23 December 2011. On 6 July 2026, the publisher reinstated both papers (Brainard and Kean 2026) and added a notice stating that they had been retracted "as a result of human error in 2011". Springer Nature thus denied that any software or "bot" had been involved in the process. The publisher nevertheless acknowledged that its records were limited and that the systems used to record integrity matters had changed. As a result, it has provided no convincing explanation of what copyright violation was originally believed to have occurred to justify retraction or how these two papers, published some seventy years earlier, were selected for review in the first place.

Therefore, we remain skeptical that the process was entirely manual. Indeed, it is difficult to imagine that, in 2011, someone at Springer independently browsed decades of digitized journal archives containing thousands of papers and happened to identify these two short papers as presenting copyright problems. Given the scale of the publisher's digitized catalogue, we still consider it more plausible that some automated workflow involving rights management or duplicate detection codes initially brought the papers to human attention. This is not to suggest that a "bot" autonomously retracted Planck's papers.

The final decision may indeed have been made by a human, but we think it more plausible that it was not an "error" but an aggressive and anachronistic interpretation of copyrights infringement, a practice that is now typical of a scientific publishing system whose first objective is to maximize profit rather than the dissemination knowledge, as had been the case for centuries.